\documentclass[prl]{revtex4}
\usepackage{amsmath}
\usepackage{color}
\usepackage{graphicx}
\usepackage{amsmath,amssymb,latexsym,epsfig,graphics,epsf,bm}
\usepackage{graphicx}
\usepackage{float}
\usepackage{placeins}
\newcommand{\br}{\mathbf r}

\newcommand{\bR}{\mathbf R}
\newcommand{\tbR}{\tilde{\mathbf R}}

\newcommand{\be}{\begin{equation}}
\newcommand{\ee}{\end{equation}}

\newcommand{\tU}{\tilde{\Phi}}
\renewcommand{\o}{\char 28}
\usepackage{color}

\begin{document}
\title{Density scaling and quasiuniversality of flow-event statistics for athermal plastic flows}
\author{Edan Lerner$^1$, Nicholas P. Bailey$^2$, and Jeppe C.~Dyre$^2$}
\affiliation{$^1$Center for Soft Matter Research, Department of Physics, New York University, New York, NY 10003 \\
$^2$DNRF Centre ``Glass and Time", IMFUFA, Department of Sciences, Roskilde University, Postbox 260, DK-4000 Roskilde, Denmark}
\date{\today}

\begin{abstract}
Athermal plastic flows were simulated for the Kob-Andersen binary Lennard-Jones system and its repulsive version in which the sign of the attractive terms is changed to a plus. Properties evaluated from simulations at different densities include the distributions of energy drops, stress drops, and strain intervals between the flow events. By reference to hidden scale invariance we show that simulations at a single density in conjunction with an equilibrium-liquid simulation at the same density allows one to predict the plastic flow-event properties at other densities. We furthermore demonstrate quasiuniversality of the flow-event statistics.
\end{abstract}

\maketitle

Solids do not flow according to the traditional definition  \cite{tabor,goodstein}. Actually, a solid {\it does} flow in principle at any finite stress, even at low temperatures \cite{hen11,kur10}. The finite-stress flow of a solid is referred to as plastic flow \cite{bri45,cot53,wer57,nem83,zai06,sch07,gre13}. Such a flow is rarely observed in practice, which is of course why solids are so useful in real life. Plastic flow is usually monitored by subjecting the solid in question to a shear deformation that increases linearly in time -- if the solid does not break, it eventually yields and flows \cite{barrat04}. In the longer run a steady state is reached in which the (shear) stress fluctuates around an average \cite{lem09}. This paper focuses on the plastic flow properties of amorphous materials, which are relevant, e.g., for metallic glasses \cite{bmg,wan12}. In this field, to date the shear-transformation-zone \cite{fal98,lan03,bou07,bou09} and the soft-glassy-rheology  \cite{sol97,sol98,sol12} theories have been the most successful frameworks to explain the rich phenomenology observed in elastoplastic flows and the yielding transition. We here supplement these theories by showing that the plastic-flow properties at one density uniquely determine those at other densities.

Plastic flows can take place even at zero temperature \cite{hen11,kur10}. For an amorphous solid like a metallic glass the zero-temperature plastic flow properties provide important information about the energy landscape of the atoms and the solid's atomic structure \cite{sam05, hen11,kur10}. For this reason athermal flows of amorphous solids have recently been studied extensively by computer simulations \cite{lem09,mal04,kar10,sal12,fio14}. Much of the focus has been on interesting finite-size scaling effects of the flow-event statistics \cite{mal04,mal06,kar10b,sal13}, and by now a good understanding has been achieved of the general nature of the self-organized-criticality observed in steady-state plastic flow \cite{lin14}. 

Plastic flows take place only as the result of large stresses. In practice, these are rarely pure shear stresses, but involve also high pressures or, occasionally, even negative pressures. It is therefore important to understand the effect of pressure on a plastic flow, and this subject has indeed  been investigated intensively for many years. Thus it is well known that the rate of plastic flow decreases dramatically with increasing pressure \cite{bridgman,spi79}, typically following an exponential function, a fact that is often rationalized in terms of the free-volume model \cite{ferry}. 

In the last decade experiments on supercooled liquids have shown that if a simple description is aimed at, the right quantity to focus on is not the pressure, but the density. Thus for a large number of glass-forming liquids in metastable equilibrium it has been demonstrated that the viscosity -- or, equivalently, the average relaxation time -- is a function of $\rho^\gamma/T$ in which $\rho$ is the density, $T$ the temperature, and $\gamma$ the so-called density-scaling exponent (not to be confused with the shear displacement that is traditionally also denoted by $\gamma$) \cite{rol05,flo11}. This so-called power-law density-scaling relation has been established convincingly by Roland, Paluch, and coworkers for many van der Waals bonded organic liquids and polymers \cite{rol05,flo11}. For large density changes the density-scaling exponent is no longer constant and a more general form of scaling takes over \cite{alb04,boh13,grz14}.

The purpose of the present paper is to show that the insight gained from the study of supercooled liquid dynamics -- that density rather than pressure is the important variable -- can be used for understanding the scaling properties of athermal plastic flows. We develop below a theory of the density dependence of the statistics of athermal plastic flow events, which makes it possible to predict the behavior at any density from the properties at one density. A first step in this direction was taken in Ref. \cite{ler09a} by assuming a constant exponent for the density scaling of the {\it average} plastic flow properties of stationary plastic flows. 

Simulations of zero-temperature flows of two simple model systems were performed in order to investigate the effect of density changes. Zero temperature (``athermal'') means the system is always at a local potential-energy minimum in the energy landscape -- thus no atom exhibits any thermal vibration. In the simulations each system was subjected to a quasistatic shear deformation, corresponding to the limit of zero strain rate. Every now and then the system exhibits a sudden plastic flow event \cite{mal04}, an avalanche that takes it from a mechanically marginally stable state, i.e., with a single zero eigenvalue of the potential-energy Hessian, to a mechanically stable state.

\begin{figure}
\centering
\includegraphics[scale = 0.47]{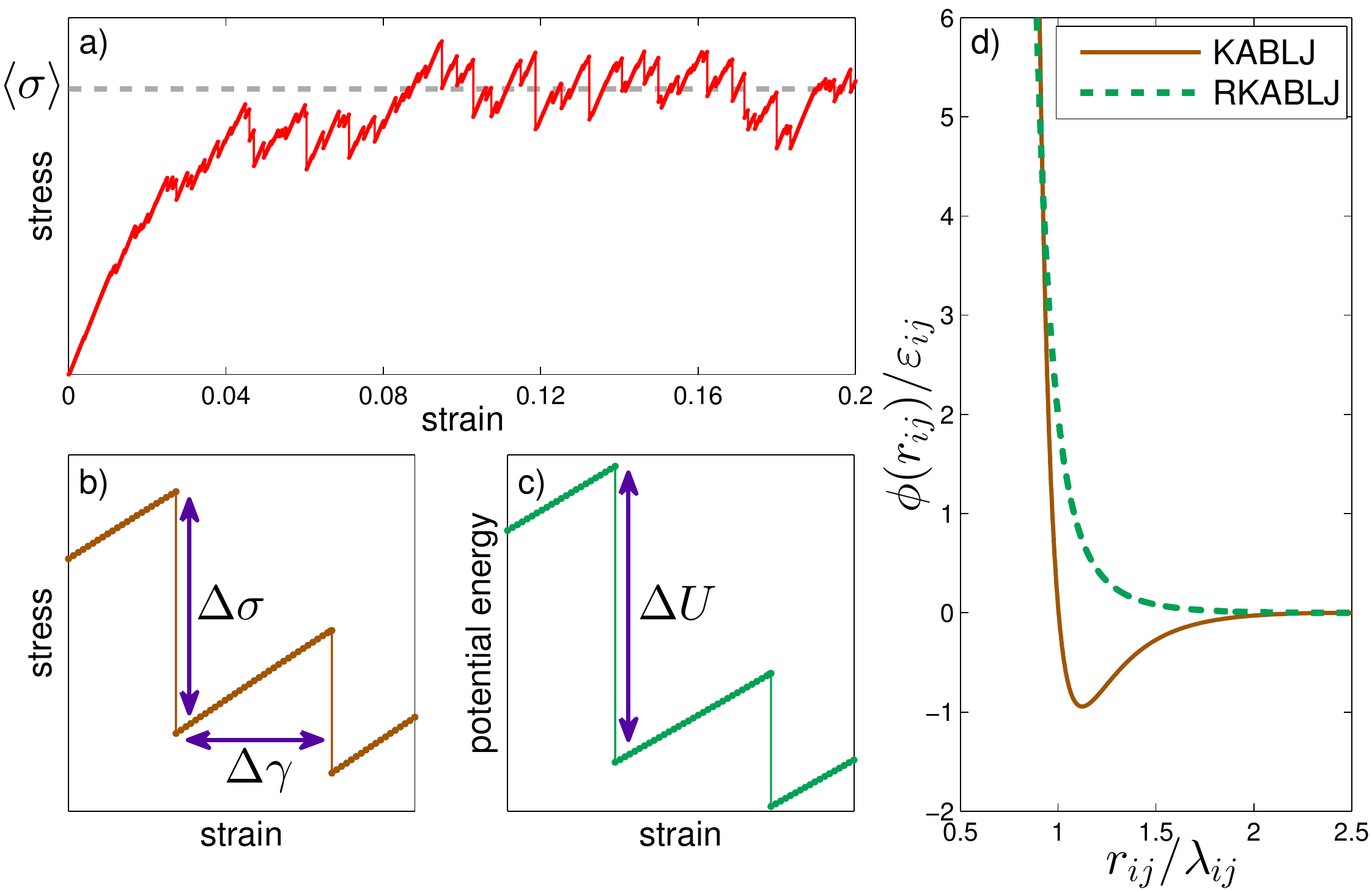}
\caption{(a) Typical stress-strain signal for a finite-size glass deformed under athermal quasistatic conditions. The dashed horizontal line indicates the steady-state flow stress $\langle\sigma\rangle$. (b) and (c) Illustration of the key observables measured in this work: the stress drops $\Delta \sigma$, the strain intervals between drops $\Delta \gamma$, and the energy drops $\Delta U$. (d) The Lennard-Jones pair potential of the Kob-Andersen binary Lennard-Jones model (KABLJ) and its purely repulsive counterpart with a plus in front of the $r^{-6}$ term (RKABLJ) as functions of the distance between particles $i$ and $j$, $r_{ij}$. The KABLJ pair potential follows the standard LJ normalization of having minimum energy $-\varepsilon_{ij}$, whereas the RKABLJ potential is normalized by requiring the energy to be $\varepsilon_{ij}$ at $r_{ij}=\lambda_{ij}$ \cite{ing12a}.}
\label{fig1}
\end{figure}

A typical stress-strain signal is displayed in Fig. 1(a). Starting from the as-quenched glass the stress initially increases linearly and few plastic flow events take place. This is the standard linear, elastic response of a solid to an imposed shear deformation. At long times a fluctuating steady state is eventually reached in which the stress saturates by increasing linearly and continuously with deformation, but dropping discontinuously whenever a flow event takes place. As mentioned, this happens when a mechanically marginally stable state is reached. At this point the potential energy drops while the system tumbles down the potential-energy landscape to a new state of mechanical equilibrium. From here the system is gradually shear deformed and reacts like a strained elastic solid, the shear stress of which increases until a new flow event takes place, the stress decreases abruptly, etc.

Following previous works in the field \cite{mal04,mal06,bailey07,ler10,sal12} we probed the statistics of the plastic flow events by determining the probability distributions of the flow events' shear-stress drops and energy changes, as well as the strain intervals between consecutive flow events (Figs. 1(b) and 1(c)). 

Specifically, we simulated two model glass-forming systems in three dimensions, each consisting of particles of two sizes. Single-component systems were not studied because, even if prepared into an amorphous state by rapid quenching from the liquid, they crystallize when subjected to plastic deformations and no steady state of glassy plastic flow can be reached \cite{bar08}. The interaction potentials simulated are the Kob-Andersen binary Lennard-Jones (KABLJ) potential of 80\% large and 20\% small particles \cite{ka1} and its repulsive counterpart (RKABLJ) \cite{ing12a}: the LJ pair potential between particles $i$ and $j$ is the usual $4\varepsilon_{ij}[(r_{ij}/\lambda_{ij})^{-12} - (r_{ij}/\lambda_{ij})^{-6}]$, the RLJ pair potential is $(\varepsilon_{ij}/2)[(r_{ij}/\lambda_{ij})^{-12} + (r_{ij}/\lambda_{ij})^{-6}]$. Particle masses do not enter into the problem.  The AQS scheme for athermal plastic-flow simulations \cite{mal99,mal04,mal06,ler10} was used. It consists of repeated applications of incremental shear-strain deformations, followed by a minimization of the potential energy using a standard nonlinear conjugate-gradient algorithm.  

Each system simulated consisted of $N=8,000$ particles in volume $V$ (density $\rho\equiv N/V$). We first deformed the system under athermal shear as described above to a strain of $\sim 1$. This was done in order to assure that steady state has been reached in which there is no memory of the initial state \cite{fio14} and the statistics of the plastic flow events is strain independent. We then started probing the statistics of the stress drops $\Delta\sigma$, the strain intervals between mechanical instabilities $\Delta\gamma$, and the potential-energy drops $\Delta U$. The algorithm used allows one to single out the drops to a high precision -- thus the instabilities are determined up to a resolution of $\sim 10^{-6}$ in strain. For each of the systems studied, statistics from about 20,000 plastic events were collected over 100 independent realizations in the steady state.

\begin{figure}
\centering
\includegraphics[scale = 0.47]{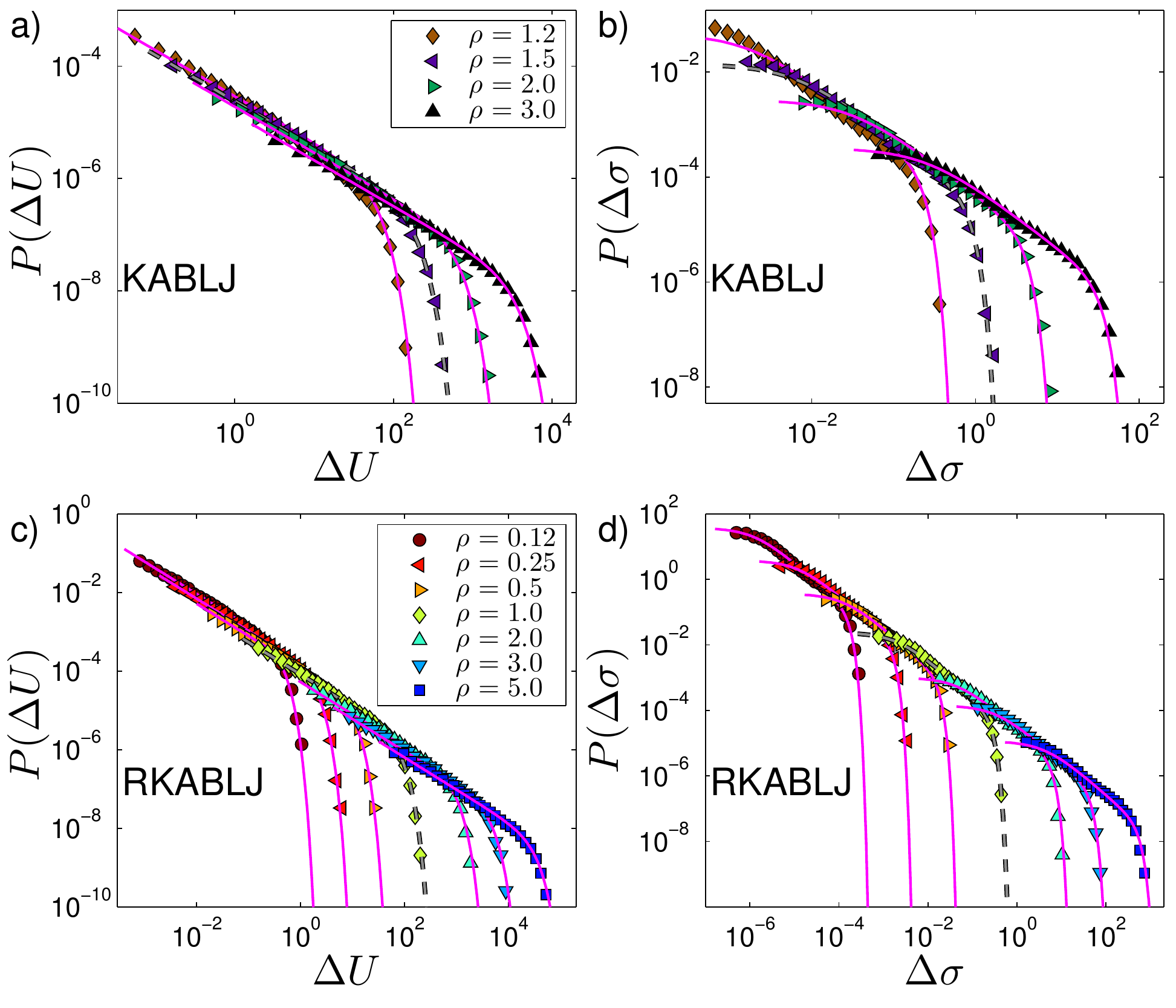}
\caption{Distributions of energy drops $P(\Delta U)$ for (a) the KABLJ and (c) the RKABLJ systems at different densities, as well as distributions of stress drops $P(\Delta \sigma)$ for (b) the KABLJ and (d) the RKABLJ systems. At the reference densities, $1.5$ for the KABLJ system and $1.0$ for the RKABLJ system, each distribution was fitted with a smooth function (dashed curves). For each system the function $h(\rho)=A\rho^4\pm B\rho^2$ was found by simulations of the equilibrium liquid at the reference densities. Based on these inputs, the distributions at the other densities are predicted uniquely from the distributions at the reference densities, see the main text (full curves).}
\label{fig2}
\end{figure}

The probability distributions of the stress drops and the potential-energy drops are shown in Fig. 2 for different densities. The lowest density simulated for the KABLJ system is $1.2$, corresponding to a slightly negative pressure. For the RKABLJ system we studied almost two decades of densities (pressure never becomes negative in a purely repulsive system). 

In order to understand the scaling properties of the probability distributions we refer to the fact that both systems have strong virial potential-energy correlations \cite{ing12a,ped08}. As shown recently, this implies that they exhibit hidden scale invariance in the sense that the potential energy $U$ as a function of the collective position vector $\bR\equiv ({\br}_1, ..., {\br}_N)$ can be written \cite{dyr13a}

\be\label{prop1m}
U(\bR)\cong h(\rho)\tU(\tbR)+g(\rho)\,.
\ee
Here $\tbR\equiv\rho^{1/3}\bR$ is so-called the reduced collective position vector, which is dimensionless, and $h(\rho)$ and $g(\rho)$ are functions of density, both of dimension energy. Equation (\ref{prop1m}) defines a  generalized, approximate scale invariance because the function $\tU(\tbR)$, which determines structure and dynamics of the equilibrium or non-equilibrium solid or liquid system, is dimensionless. In particular, $\tU(\tbR)$ does not involve any of the parameters of the pair potentials $\varepsilon_{ij}$ and $\lambda_{ij}$. 

According to Eq. (\ref{prop1m}) a change of density to a good approximation results in a linear, affine overall scaling of the high-dimensional potential-energy surface. For systems in thermal equilibrium this can be compensated by adjusting the temperature. This gives rise to the so-called isomorphs \cite{IV,dyr13a}, which are curves in the thermodynamic phase diagram given by $h(\rho)/T={\rm Const.}$ along which structure and dynamics of equilibrium systems are invariant to a good approximation. 

Whereas the function $h(\rho)$ is important for determining the energy scale for the dynamics at a given density, $g(\rho)$ merely provides an additive constant to the potential energy and plays no role for structure and dynamics (this function contributes, of course, to the pressure and its density dependence and thus to the equation of state). Since $\tU(\tbR)$ is density independent, by dimensional analysis \cite{jhj13} Eq. (\ref{prop1m}) implies that for different densities the distribution of flow-event potential-energy drops, $p(\Delta U)$, is of the form $p(\Delta U)=F_U\big(\Delta U/h(\rho)\big)/h(\rho)$ in which the function $F_U$ is density independent. Likewise by dimensional analysis, the distribution of shear-stress drops must be of the form $p(\Delta\sigma)=F_\sigma\big(\Delta\sigma/(\rho h(\rho))\big)/(\rho h(\rho))$. In order to test these predictions we proceeded as follows (Fig. 2). 

First, the function $h(\rho)$ was determined by making use of the fact \cite{ing12a,boh12} that for a system with pair potentials of the form $v(r)=\sum_n \varepsilon_n (r/\sigma)^{-n}$ one has $h(\rho)=\sum_n \alpha_n \varepsilon_n(\rho \sigma^3)^{n/3}$ for suitable constants $\alpha_n$. In the present case, each of the two $h(\rho)$ functions has two parameters, $h(\rho)=A\rho^4\pm B\rho^2$, but since the overall scaling of $h(\rho)$ is undetermined, only a single parameter needs to be fixed, e.g., $B/A$. This was done by simulations of the equilibrium liquid at a reference density by equating the expression for the density-scaling exponent, $\gamma=d\ln h/d\ln\rho$ \cite{ing12a,boh12,dyr13a}, to its canonical-ensemble fluctuation expression $\gamma={\langle\Delta W \Delta U\rangle}/{\langle (\Delta U)^2\rangle}$ calculated from an $NVT$  simulation \cite{IV}. As reference densities we chose $\rho=1.5$ for the KABLJ system and $\rho=1.0$ for the RKABLJ system. The density-scaling exponent was determined at these densities at $T=2.0$ for the KABLJ system ($\gamma=4.65$, leading to also $B/A=0.55$) and at $T=1.0$ for the RKABLJ system ($\gamma=3.29$, leading also to $B/A=0.55$). Once $h(\rho)$ has been determined, the plastic flow-event statistics at one density -- marked by the green dashed curves in Fig. 2 -- uniquely predict those at the other densities (full curves). 

As a further investigation of the scaling properties predicted from hidden scale invariance Fig. 3 shows the average stress $\langle\sigma\rangle$, the average stress drop $\langle\Delta\sigma\rangle$, and the average potential-energy drop $\langle\Delta U\rangle$, as functions of density for the two systems. The full curves are the predictions from dimensional analysis: $\langle\sigma\rangle\propto\rho h(\rho)$, $\langle\Delta\sigma\rangle\propto\rho h(\rho)$, and $\langle U\rangle\propto h(\rho)$. For each of the three quantities the proportionality constant was found by fitting at the reference density.

The scaling properties identified above allow one to compare different systems' flow-event statistics. Figure 4 shows data for the simulations, supplemented with data for the average shear displacement between flow events. The figure also includes data for an inverse-power-law (IPL) binary system with pair potential $v(r)\propto r^{-10}$, for which the scaling is exact and trivial. These three quite different systems (compare Fig. 1(d)) have very similar flow-event properties. We interpret this as reflecting the quasiuniversality \cite{ros77,ros80c} previously discussed only for single-component liquid model systems in thermal equilibrium (see, e.g., Ref. \cite{dyr13} and its references). In the hidden-scale-invariance language, quasiuniversality at thermal equilibrium implies that the function $\tU(\tbR)$ in Eq. (\ref{prop1m}) is quasiuniversal, which implies quasiuniversality of the flow-event statistics. Although in the different context of universality upon approach to the jamming transition, we note that Arevalo and Ciamarra recently also reported quasiuniversality of flow-event properties \cite{are13}.

In summary, the hidden scale invariance property implies that from simulations at a single density one can predict the plastic flow-event properties at arbitrary densities. This was confirmed by simulations for both the KABLJ and the RKABLJ systems. Moreover, the simulations show a quasiuniversality for the plastic flow properties, the generality of which remains to be investigated.

\begin{figure}
\centering
\includegraphics[scale = 0.47]{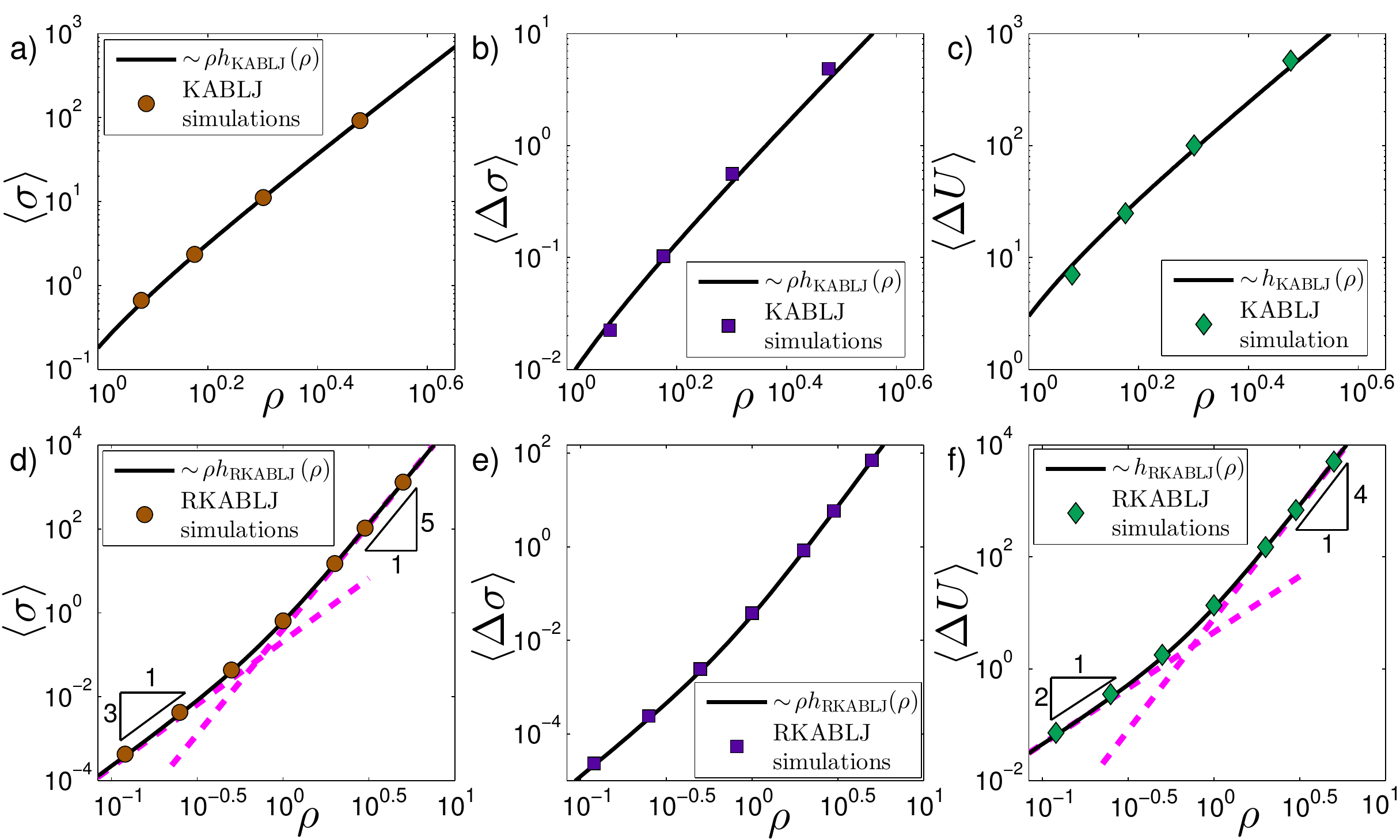}
\caption{Mean steady-flow observables as functions of density for the KABLJ (top panels) and RKABLJ (bottom panels) systems. The continuous curves are proportional to $h(\rho)$ or $\rho h(\rho)$ for quantities having units of energy or stress, respectively. The dashed lines indicate the prediction of power-law functions $h(\rho)$ corresponding to constant density-scaling exponents.}
\label{fig3}
\end{figure}

\begin{figure}
\centering
\includegraphics[scale = 0.47]{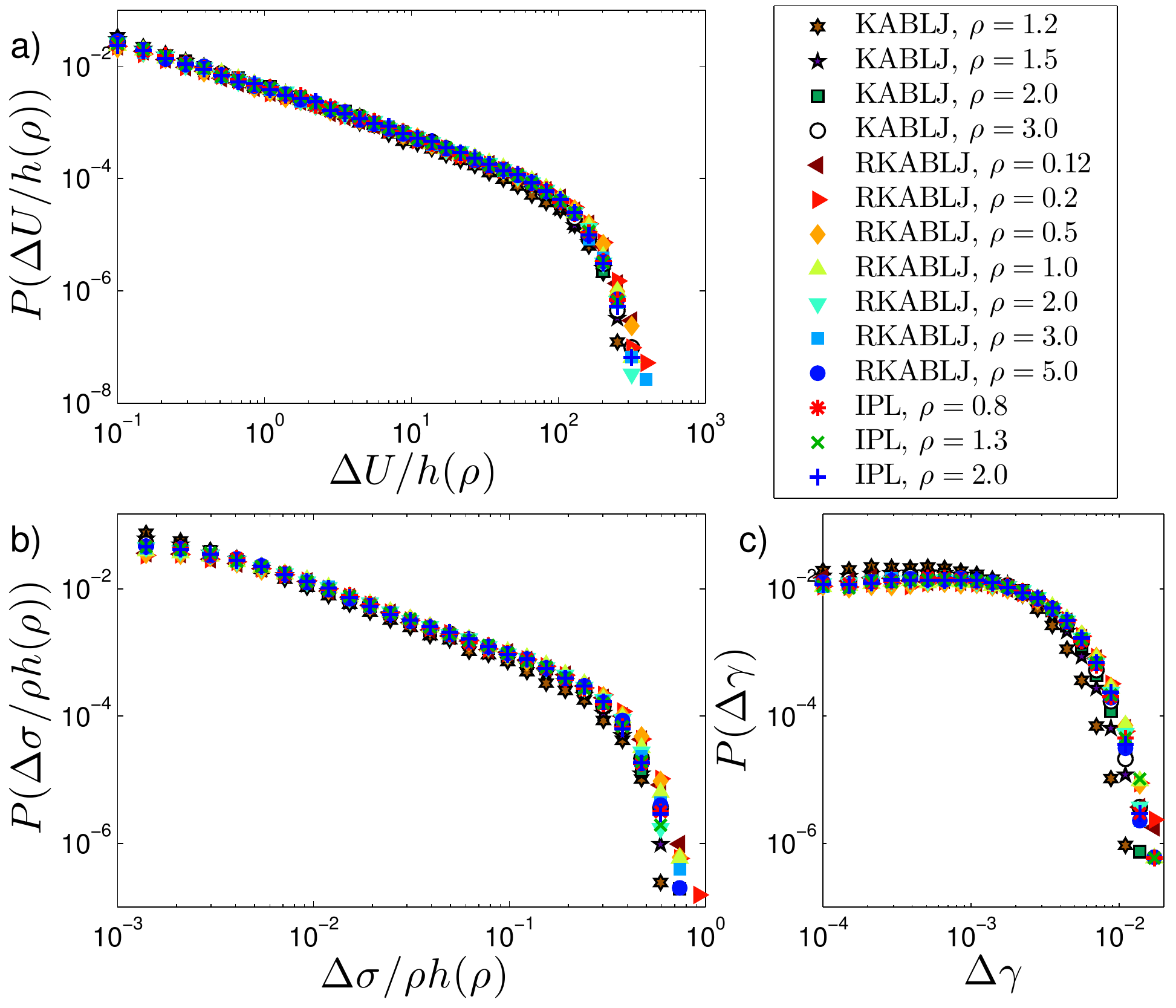}
\caption{Distributions of energy drops ${\Delta U}/{h(\rho)}$, stress drops ${\Delta \sigma}/({\rho h(\rho)})$, and strain intervals $\Delta\gamma$ for the KABLJ, RKABLJ, and $n=10$ inverse power law (IPL) systems. This figure demonstrates quasiuniversality of the plastic flow-event properties.}
\label{fig4}
\end{figure}

{\acknowledgments 
EL acknowledges the MRSEC Program of the NSF DMR-0820341 for partial funding. 
The centre for viscous liquid dynamics ``Glass and Time'' is sponsored by the Danish National Research Foundation's grant DNRF61. 
}

 \end{document}